\documentclass{article}
\usepackage{amsmath,graphicx,mlspconf}
\usepackage{amssymb}  
\usepackage{verbatim}
\usepackage{cite}
\usepackage{graphicx}
\usepackage{textcomp}
\usepackage{subfigure}

\usepackage{graphicx}
\usepackage{textcomp}
\usepackage{float}  

\def\BibTeX{{\rm B\kern-.05em{\sc i\kern-.025em b}\kern-.08em
    T\kern-.1667em\lower.7ex\hbox{E}\kern-.125emX}}
\usepackage{comment}
\usepackage{subfigure}
\usepackage[ruled,vlined]{algorithm2e}
\SetKwInput{Input}{Input}
\SetKwInput{Output}{Output}
\SetKw{To}{ to }
\usepackage{xcolor}
\usepackage{hyperref}
\copyrightnotice{U.S.\ Government work not protected by U.S.\ copyright}

\copyrightnotice{979-8-3503-2411-2/25/\$31.00 {\copyright}2025 Crown}

\copyrightnotice{979-8-3503-2411-2/25/\$31.00 {\copyright}2025 European Union}

\copyrightnotice{979-8-3503-2411-2/25/\$31.00 {\copyright}2025 IEEE}

\toappear{2025 IEEE International Workshop on Machine Learning for Signal Processing, Aug.\ 31-- Sep.\ 3, 2025, Istanbul, Turkey}

\title{Hypergraph Overlapping Community Detection for  brain networks}
\name{Duc Vu and Selin Aviyente}
\address{Department of Electrical and Computer Engineering, Michigan State University, East Lansing, MI}

\begin{document}

\maketitle

\begin{abstract}
Functional magnetic resonance imaging (fMRI) has been commonly used to construct functional connectivity networks (FCNs) of the human brain. TFCNs are primarily limited to quantifying pairwise relationships between ROIs ignoring higher order dependencies between multiple brain regions. Recently, hypergraph construction methods from fMRI time series data have been proposed to characterize the high-order relations among multiple ROIs. While there have been multiple methods for constructing hypergraphs from fMRI time series, the question of how to characterize the topology  of these hypergraphs remains open. In this paper, we make two key contributions to the field of community detection in brain hypernetworks. First, we construct a hypergraph for each subject capturing high order dependencies between regions. Second, we introduce a spectral clustering based approach on hypergraphs to detect overlapping community structure. Finally, the proposed method is implemented to detect the consensus community structure across multiple subjects. The proposed method is applied to resting state fMRI data from Human Connectome Project to summarize the overlapping community structure across a group of healthy young adults.
\end{abstract}
\section{Introduction}
Functional magnetic resonance imaging (fMRI) is a commonly used neuroimaging modality to study the brain organization and cognitive function. In particular, fMRI
based functional connectivity networks (FCNs) have provided important insights into brain organization for healthy and disease populations as well as understanding of 
individual variability in brain activation patterns during
 cognitive tasks \cite{van2010exploring}. Conventional functional connectivity (FC) analysis measures the Pearson's correlation
between regions-of-interest (ROIs) and is thus limited to quantifying pairwise interactions in the brain. However, high-order relationships, i.e., the interactions between one ROI and multiple ROIs, play an important role in understanding  neural processes and the pathological underpinnings of diseases \cite{li2023aberrant,gatica2021high}. Therefore, in recent years, hypergraphs have been employed to model the brain network \cite{xiao2019multi,xiao2020hypergraph}. 

The hypergraph model based brain network, i.e., hypernetwork,  is composed of nodes and
hyperedges, where nodes represent ROIs and hyperedges can
connect any number of ROIs. Existing work on brain hypernetworks focus mostly on the construction of the hypergraphs from neuroimaging data \cite{xiao2020hypergraph}. In particular, they construct unweighted or weighted hypergraphs where the weights may be learned from the data and represent the hypergraph using either the incidence matrix or a tensor. The resulting representations are used in subsequent supervised learning tasks. However, in order to better understand the advantages of capturing high-order relations in brain networks, one needs to reduce the hypernetworks into a small number of interpretable components, e.g., communities.


Community organization is a crucial property of brain networks, responsible for shaping communication processes and underpinning brain functioning. Existing community detection methods for brain networks primarily focus on detecting nonoverlapping communities from FCNs constructed by computing pairwise relationships between ROIs \cite{garcia2018applications}. Thus, the resulting communities cannot capture high-order relationships and the possible dependencies between individual communities.  In this paper, we address this issue by introducing an overlapping community detection method for brain hypernetworks. The proposed method first constructs hypergraphs corresponding to each subject and applies a spectral clustering method on the line graphs corresponding to individual hypergraphs. The community memberships across subjects are then aggregated through an association matrix. Finally, the group (consensus) community structure is obtained by applying spectral clustering to the group-level hypergraph. 

\section{Background}
\subsection{Hypergraphs}
\begin{figure}
    \centering
    \includegraphics[width=1\linewidth]{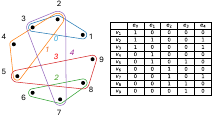}
    \caption{A hypergraph $\mathcal{G}$ with 9 nodes and 5 hyperedges. The hypergraph is given on the left and the corresponding incidence matrix $\mathbf{H}$ is given on the right.}
    \label{fig:F1}
\end{figure}

A hypergraph with $N$ vertices and $K$ hyperedges is denoted as $\mathcal{G} = (V, E)$, where $V$ is the set of nodes and $E = \{e_{1}, e_{2}, \ldots, e_{K}\}$ is the set of hyperedges. The diagonal matrix $\mathbf{W} = \operatorname{diag}\{w(e_{1}), w(e_{2}), \ldots, w(e_{K})\}$ represents the weights of the hyperedges. The hypergraph can be defined by an incidence matrix $\mathbf{H} \in \mathbb{R}^{N \times K}$, where $H_{i,j} = 1$ if $v_{i} \in e_{j}$, and $H_{i,j} = 0$ otherwise (see Fig. \ref{fig:F1}). 
In our work, we only consider $e$-uniform hypergraphs, meaning that all hyperedges of $\mathbf{H}$ have the same cardinality $e$.
 The weighted line graph of a hypergraph is defined as \( \Gamma(\mathbf{H}) \in \mathbb{R}^{K \times K} \) where each entry is the Jaccard similarity between the hyperedges \( e_i,e_j \in E\)\cite{lotito2024hyperlink}:
\begin{equation}
\Gamma(\mathbf{H})_{ij} = J(e_i, e_j) = \frac{|e_i \cap e_j|}{|e_i \cup e_j|}.
\label{eq:line-graph-jaccard}
\end{equation}

\subsection{Spectral Clustering}

Given $\mathcal{G} = (V, E, \mathbf{A})$, where $V$ is the set of nodes with $|V| = N$, $E$ is the set of edges, and $\mathbf{A} \in \mathbb{R}^{N \times N}$ is the weighted, undirected adjacency matrix, the goal of clustering is to find a partition of the graph into $K$ communities $\mathcal{C}=\{\mathcal{C}_{1},\ldots,\mathcal{C}_{K}\}$. The graph can also be represented using its normalized Laplacian matrix, defined as 
$\mathbf{L}=\mathbf{D}^{-1/2}(\mathbf{D-A})\mathbf{D}^{-1/2}$,
where $\mathbf{D}$ is the diagonal degree matrix with $D_{ii}=\sum_{j} A_{ij}$.
Community detection can be performed by minimizing the normalized cut, which is formulated through spectral clustering as follows \cite{vonluxburg2007tutorialspectralclustering}:
\begin{equation}
\underset{\mathbf{U},\, \mathbf{U}^\top \mathbf{U} = \mathbf{I}}{\text{minimize}} \; \text{tr}(\mathbf{U}^\top \mathbf{L} \mathbf{U}),
\label{eq:SC}
\end{equation}
\noindent where $\mathbf{U} \in \mathbb{R}^{N \times K}$ is an embedding matrix, and $K$ is the number of communities. The solution to the trace minimization problem is the matrix $\mathbf{U}$ whose columns are the $K$ eigenvectors of $\mathbf{L}$ corresponding to the smallest eigenvalues. The final partition is obtained by applying $k$-means clustering to the rows of $\mathbf{U}$.

\subsection{Overlapping Community Detection Using Edges}
In the case of overlapping community detection, the set of communities is called  \emph{cover} \( \mathcal{C} = \{\mathcal{C}_1, \mathcal{C}_2, \ldots, \mathcal{C}_K\} \), in which a node may belong to one or more communities \cite{10.1145/2501654.2501657}. Each node \( i \) is associated with a membership vector \( \mathbf{y}_i = [a_{i1}, a_{i2}, \ldots, a_{iK}] \), where \( a_{ic} \in \{0,1\}\) is the node membership indicator, i.e., \( a_{ic} = 1 \) when node \( i \) is in community \( c \), and \( 0 \) otherwise. 

To detect the community of edges,  any clustering algorithm can be applied to the Jaccard similarity matrix between edges as it behaves similarly to an adjacency matrix for hyperedges \cite {Ahn_2010, lotito2024hyperlink}. Nodes inherit the community membership of the hyperedges they participate in \cite{lotito2024hyperlink}. The node community membership matrix \( \mathbf{Y} \in \mathbb{R}^{N \times K} \) is then defined as:
\begin{equation}
Y_{ij} = 
\begin{cases}
1 & \text{if node } i \text{ is in community } j, \\
0 & \text{otherwise}.
\end{cases}
\label{eq:membership}
\end{equation}

\section{Methodology}
\subsection{Hypergraph Learning Using FCN Construction}

Let \( \mathbf{F} = \{\mathbf{f}_1, \mathbf{f}_2, \ldots, \mathbf{f}_N\} \) denote the fMRI time series for a subject with \( N \) regions of interest (ROIs), where \( \mathbf{f}_i \in \mathbb{R}^P \) is the time series of the \( i \)-th ROI, and \( P \) is the number of time points. We normalize \( \mathbf{F} \) such that each row has zero mean and unit Euclidean norm. This standardization allows the use of consistent hyperparameter values across subjects.

To characterize higher-order interactions among multiple ROIs, we first construct the hyperedge weight vectors \( \mathbf{w}_i \) and compute the corresponding incidence matrix \( \mathbf{H} \) using sparse linear regression (i.e., Lasso) \cite{guo2018resting}. Specifically, for each ROI’s time series \( \mathbf{f}_i \), we learn the weights \( \mathbf{w}_i \in \mathbb{R}^{N-1} \) that model the dependency between ROIs as follows:

\begin{equation}
\min_{\mathbf{w}_i} \frac{1}{2} \left\| \mathbf{f}_i - \mathbf{B}_i \mathbf{w}_i \right\|_2^2 + \lambda \left\| \mathbf{w}_i \right\|,
\label{eq:sparse_regression}
\end{equation}

\noindent where \( \mathbf{B}_i = \{\mathbf{f}_1, \mathbf{f}_2, \ldots, \mathbf{f}_{i-1}, \mathbf{f}_{i+1}, \ldots, \mathbf{f}_N\} \) is the time series of all the ROIs except the \( i \)-th one. The vector \( \mathbf{w}_i \in \mathbb{R}^{N-1} \) is the representation coefficient vector for the \( i \)-th ROI in terms of the other ROIs. \( \lambda > 0 \) is a regularization parameter that controls the sparsity of \( \mathbf{w}_i \). 

Based on the sparse representation in Eq. \ref{eq:sparse_regression},  a hyperedge consists of a centroid ROI (the \( i \)-th ROI) and other ROIs whose corresponding coefficients in the weight vector \( \mathbf{w}_i \) are positive. Negative coefficients are ignored, as they indicate a negative influence on the centroid ROI \cite{guo2018resting}. This process filters out insignificant interactions between the centroid and the other ROIs while preserving the most important interactions. In this manner, we obtain \( N \) hyperedges for each ROI.

We then define \( \tilde{\mathbf{w}}_i \) by setting the \( i \)-th element of \( \mathbf{w}_i \) to zero, resulting in \[
\tilde{\mathbf{w}}_i = (w_{i1}, w_{i2}, \dots, w_{i,i-1}, 0, w_{i,i+1}, \dots, w_{in}) \in \mathbb{R}^N.
\]
To obtain hyperedges with cardinality $e$, we select the ROIs corresponding to the largest  \( e-1 \) elements of \( \tilde{\mathbf{w}}_i \) along with the \( i \)-th ROI to form the \( i \)-th hyperedge. This ensures that the centroid ROI and the ROIs with highest interactions with that centroid are included in the same hyperedge. Using these hyperedges, we construct the incidence matrix \( \mathbf{H} \in \mathbb{R}^{N \times N} \).
\noindent Since \( \mathbf{H} \) is built from \( \mathbf{w}_i \)'s, it becomes sparser as \( \lambda \) increases and denser as \( \lambda \) decreases.

\subsection{Spectral Clustering on Line Graph}

 To detect overlapping communities, we first construct the weighted line graph of a hypergraph \(\Gamma(\mathbf{H}) \in \mathbb{R}^{N \times N} \) using the pairwise Jaccard similarity of hyperedges (\ref{eq:line-graph-jaccard}).  We then apply spectral clustering to detect the communities of hyperedges. 
The number of communities is determined using the eigengap rule, where the \( K \)-th largest eigengap of the normalized Laplacian, $\mathbf{D}^{-1/2} ( \mathbf{D} - \Gamma(\mathbf{H}) ) \mathbf{D}^{-1/2}$
,  of the line graph is considered the optimal number of communities. With the hyperedge communities, we construct the node community membership matrix \( \mathbf{Y} \in \mathbb{R}^{N \times K} \)(\ref{eq:membership}). The corresponding subject-level community structure matrix \( \mathbf{M} \in \mathbb{R}^{N \times N} \) is: 

\begin{equation}
M_{ij} = 
\begin{cases}
1, & \text{if } v_i \text{ and } v_j \text{ are in the same community} \\
0, & \text{otherwise}.
\end{cases}
\label{eq:M}
\end{equation}


\subsection{Association Matrix and Consensus Community}


Once we obtain the subject-level community structure matrix \( M^{sr} \) for each subject \( s \in \{1, \dots, S\} \) and run \( r \in \{1, \dots, R\} \), where \( S \) is the number of subjects and \( R \) is the number of runs, we compute the association matrix:

\begin{equation}
\mathbf{A} = \frac{1}{SR} \sum_{s=1}^{S} \sum_{r=1}^{R} \mathbf{M}^{sr}.
\label{eq:associate}
\end{equation}
This can be seen as the group level adjacency matrix, as it reflects how often two nodes appear in the same community, indicating their connectivity. Treating each node in \( \mathbf{A} \) as a centroid, we obtain group-level hyperedges by selecting the top-order \( e \) nodes for each node to form a hyperedge. We construct the incidence matrix \( \mathbf{H}_{group} \) from the hyperedges and repeat the process outlined above, i.e., calculate the Jaccard similarity of the hyperedges to construct the line graph, determine the number of communities using the eigengap, and perform spectral clustering to obtain the consensus communities \( \mathbf{Y}_{group} \) in the node domain. The steps of the algorithm are summarized in Algorithm \ref{alg} \footnote{The code is available at \url{https://github.com/tienducvunguyen/HyperGraphOverlapping.git}.}. 

\begin{algorithm}[ht]
\caption{Overlapping Community Detection}
\label{alg}
\LinesNumbered
\Input{ fMRI time series matrix $\mathbf{F} = (\mathbf{f}_1, \ldots, \mathbf{f}_N) \in \mathbb{R}^{N \times P}$ for $S$ subjects with $K$ runs.}
\Output{\( \mathbf{Y}_{group} \)}

Normalize each row of $\mathbf{F}$ to zero mean and unit norm\;

\For{$s\in \{1,2,\ldots,S\}$ \text{and} $r\in \{1,2,\ldots,R\}$}{
    \For{each ROI $i = 1$ \KwTo $N$}{
        Form $\mathbf{B}_i \in R^{N -1 \times P} \gets \mathbf{F} \setminus \mathbf{f}_i$\;
        
        Solve sparse regression (\ref{eq:sparse_regression});
        
        Retain top $e - 1$ largest positive entries in $\mathbf{w}_i$\ along with i-th entry;
    }
    
    Construct incidence matrix $\mathbf{H}$ using retained weights\;
    
    Compute line graph $\Gamma(\mathbf{H})$(\ref{eq:line-graph-jaccard});
    
    Perform spectral clustering on $\Gamma(\mathbf{H})$(\ref{eq:SC});
    
    Get node community membership matrix $\mathbf{Y}$\ (\ref{eq:membership});
    
    Form community structure matrix $\mathbf{M}$ (\ref{eq:M};
}

Compute group-level association matrix $\mathbf{A}$ (\ref{eq:associate});

    \For{each ROI $i = 1$ \KwTo $N$}{
    Select top $e - 1$ neighbors in $\mathbf{A}$ along with ROI i to form a hyperedge\;
}

Repeat lines 9, 10, 11 on group-level hypergraph to obtain consensus communities $\mathbf{Y_{group}}$\;
\end{algorithm}

\section{Result}

\subsection{Human Connectome Project Data}
The proposed method is applied on functional neuroimaging data from 56 subjects collected as part of the Human Connectome Project (HCP)\footnote{\url{db.humanconnectome.org}}. The subjects are selected from HCP's healthy young adult study (HCP 900) and include 34 females and 22 males, in the age range 26-35. 
Data acquisition is performed using a Siemens 3T Skyra with a 32-channel head coil \cite{van2013wu}. The scanning protocol includes high-resolution T1-weighted scans (256 slices, $0.7 \, \text{mm}^{3}$ isotropic resolution, $\text{TE} = 2.14 \, \text{ms}$, $\text{TR} = 2400 \, \text{ms}$, $\text{TI} = 1000 \, \text{ms}$, flip angle $=8^\circ$, $\text{FOV} = 224 \times 224 \, \text{mm}^2$, $\text{BW} = 210 \, \text{Hz/px}$, $\text{iPAT} = 2$) \cite{glasser2013minimal}. Functional scans were collected using a multi-band sequence with MB factor $8$, isotropic $2 \, \text{mm}^3$ voxels, $\text{TE} = 33 \, \text{ms}$, $\text{TR} = 720 \, \text{ms}$, flip angle $= 52^\circ$, $\text{FOV} = 208 \times 180 \, \text{mm}^2 \, (\text{RO} \times \text{PE})$, 72 slices, $2.0 \, \text{mm}$ isotropic resolution, $\text{BW} = 290 \, \text{Hz/px}$, echo spacing $= 0.58 \, \text{ms}$ \cite{glasser2013minimal}. One hour of resting state data was acquired per subject in 15-minute intervals over two separate sessions with eyes open and fixation on a crosshair. Within each session, oblique axial acquisitions alternated between phase encoding in a right-to-left (RL) direction in one run and left-to-right (LR) in the other run. Minimally ICA-FIX cleaned resting-state fMRI (rs-fMRI) are used in the following analysis. Following this denoising step, functional volumes are
spatially smoothed with a Gaussian kernel (5 mm full-width
at half-maximum). The first 10 volumes are discarded so that
the fMRI signal achieves steady-state magnetization, resulting
in P = 1190 time points. Voxel fMRI time courses are
detrended and band-pass filtered [0.01 - 0.15] Hz to improve
the signal-to-noise ratio for typical resting-state fluctuations.
Finally, Glasser’s multimodal parcellation resliced to fMRI resolution
is used to parcellate fMRI volumes and compute regionally
averaged fMRI signals. These were z-scored and stored in an
$N \times P$ matrix.
\subsection{Community Detection}
Our framework is applied to HCP data, $\mathbf{F} \in \mathbb{R}^{N \times P}$, from two sessions. We set \( \lambda = 0.01 \) in the sparse linear regression problem \cite{xiao2019multi}. The hyperedge order is set to be minimum such that the line graphs constructed using Jaccard similarity are fully connected across subjects. For all subjects, we found the hyperedge order to be \( e = 4 \) using this criterion. For the group-level line graph, the hyperedge order is set to \( e = 6 \).

\begin{figure}
    \centering
    \includegraphics[width=0.8\linewidth]{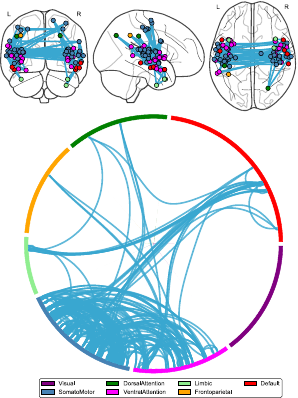}
    \caption{The hyperedges of the first community detected for the first subject. Each hyperedge is  displayed as the pairwise edges (blue) between all the nodes in the hyperedge. The chord diagram and the sagittal cut view are shown.}
    \label{fig:hyperedge_chord}
\end{figure}

\begin{figure}[t]
    \centering
    \includegraphics[width=1\linewidth]{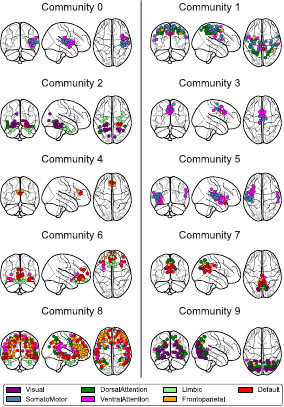}
    \caption{10 communities detected from the group-level association matrix after processing every subjects' fMRI data. Most of the detected communities  are symmetric or is symmetric with another community.}
    \label{fig:all_communities}
\end{figure}

As shown in Fig.~\ref{fig:all_communities},  \(10\) group-level communities are identified. The communities are symmetric and highly localized. We can see resemblance to important ICA-identified resting-state connectivity patterns~\cite{beckmann2005investigations}. For instance, community~9 and 7 relate to the Lateral Visual Cortical area and Visuo-Spatial System, respectively. Furthermore, detected community can be combined to form familiar patterns, such as community~0 and 5 combined correspond to the Auditory System, community 1 and 3 combined is similar to the Sensory Motor system, and community~4 and 6 combined is similar to the Executive Control area. The algorithm was able to detect overlapping areas of functions, such as Executive Control which provides bias signals to other areas to implement cognitive control, and the Visuo-Spatial system which serves to engage in visuo-spatial information around us.

\begin{figure}
    \centering
    \includegraphics[width=1\linewidth]{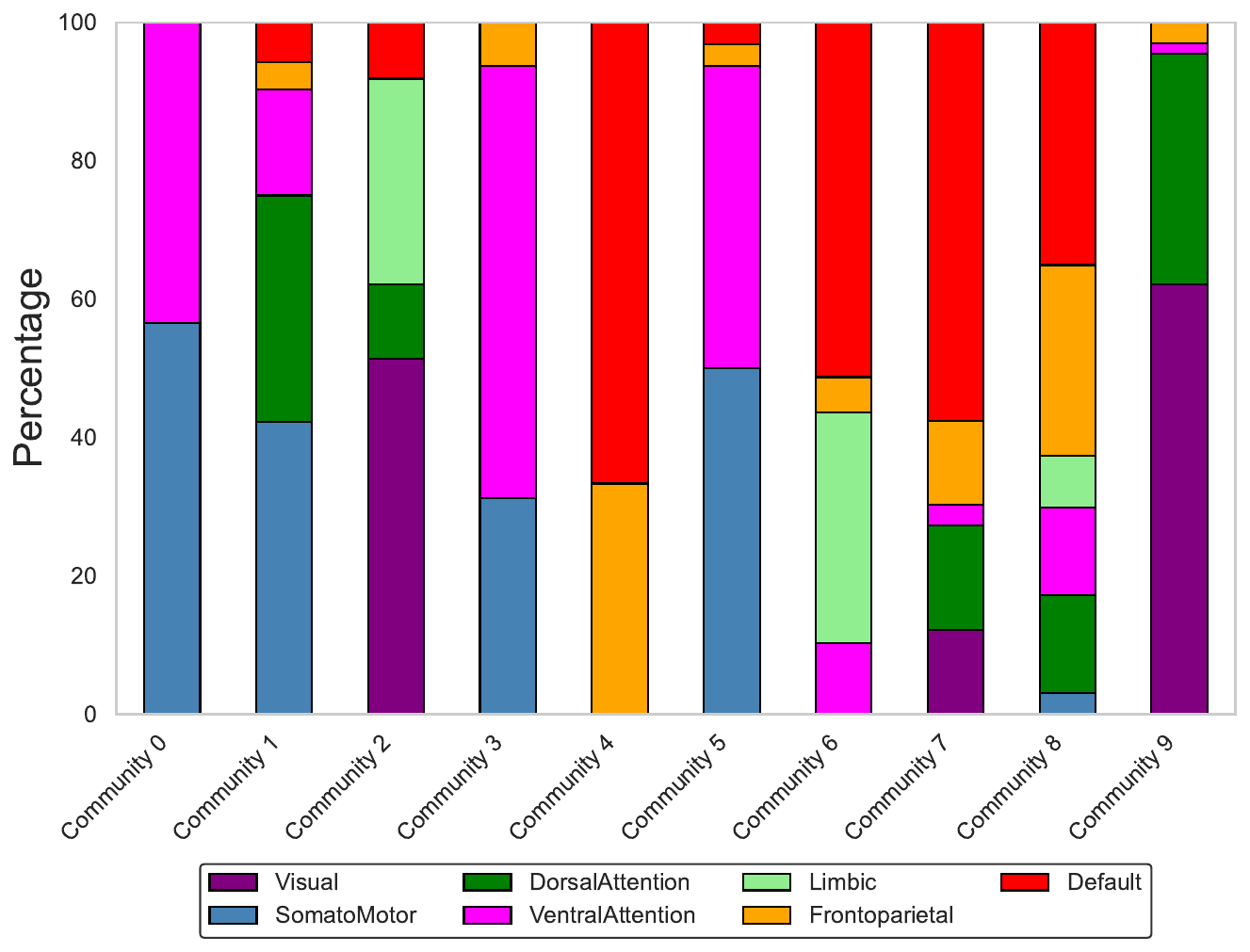}
    \caption{The percentage of nodes within each community that belongs to a single Yeo region. }
    \label{fig:percentage}
\end{figure}

We calculated the percentage of Yeo network regions in each of the communities (Fig.~\ref{fig:percentage}), and observe that many communities are focused on one or two networks, such as communities~0, 3, and 5 with only SomatoMotor and Ventral Attention networks, and community~9 with Dorsal Attention and SomatoMotor networks. Communities~4, 6, and 7 are mostly dominated by Default Mode and FrontoParietal networks. Along with their dense connections (Fig.~\ref{fig:hyperedge_chord}), this means that the detected communities are functional combinations of different Yeo regions.

\begin{figure}
    \centering
    \includegraphics[width=1\linewidth]{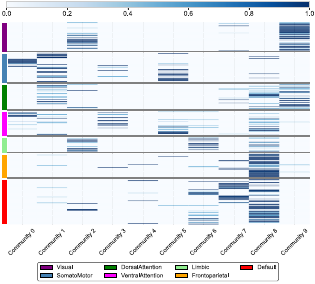}
    \caption{Heatmap showing the membership strength of each node across communities. Each row represents a node, grouped by its Yeo network affiliation (indicated by the multi-colored bar). Columns correspond to communities, with cell values indicating the node's relative membership strength. Nodes within multiple communities have their membership strength normalized so  that the sum is equal to 1.}
    \label{fig:community_assignment}
\end{figure}

The size of the communities varies, with Community 4 containing only 6 nodes, while Community 8 containing 134 nodes. Community 8 actually comprises over 80\% of the total frontal-parietal nodes and 60\% of the default mode network nodes (Fig.~\ref{fig:community_assignment}). This community represents the strong connection between the default mode network and the frontal-parietal network as these networks are primarily responsible for cognitive functions \cite{hearne2016functional} at rest.

According to Fig. \ref{fig:community_assignment}, Community 8's nodes have significant overlap with Community 1, indicating a connection between cognitive functions and sensorimotor functions. Similarly, Community 6 shows substantial overlap with community 8, suggesting a connection between the Executive Control area and cognitive functions. On the other hand, SomatoMotor network had very few overlapping nodes, indicating isolated networks\cite{yeo2014estimates}.


To quantify the consistency of the detected communities across subjects, we assigned a score to each community (Table~\ref{tab:community_cscore_stats}). Let \( \mathbf{Q} \in \mathbb{R}^{R \times S \times K} \) with \( R \) being the number of runs, \( S \) being the number of subjects, and \( K \) being the number of communities. \( Q_{C_i}^{\mathrm{sr}} \) is equal to the number of hyperedges entirely within community \({C_i}\) for subject $s$ in run $r$, divided by the number of hyperedges associated with that community. A hyperedge is within a community only if all its nodes are members, and associated with a community if at least one of its nodes is a member. Table~\ref{tab:community_cscore_stats} reports the mean and standard deviation of \( {Q}_{C_i} \) across all runs and subjects. To verify the statistical significance of these values, we calculated the $p$-value for each \( {Q}_{C_i} \)  by comparing with \( \mathbf{Q}_{random} \), which is obtained by keeping the number of nodes in each community at a constant and choosing the nodes within each community at random  \( 100 \) times. For all communities, the computed score was significant at  $p<0.001$. 



\begin{table}[h!]
\centering
\scriptsize
\resizebox{0.8\linewidth}{!}{%
\begin{tabular}{c|c|c}
Community & \( {Q}_{C_i} \) & \( {Q}_{C_\text{i-random}} \)\\
\hline
\hline
0 & 0.3351 ± 0.0646 & 0.0001 ± 0.0008 \\
1 & 0.2726 ± 0.0428 & 0.0008 ± 0.0023 \\
2 & 0.1974 ± 0.0543 & 0.0003 ± 0.0017 \\
3 & 0.2010 ± 0.0634 & 0.0000 ± 0.0005 \\
4 & 0.0956 ± 0.0717 & 0.0000 ± 0.0000 \\
5 & 0.2557 ± 0.0573 & 0.0002 ± 0.0012 \\
6 & 0.1865 ± 0.0383 & 0.0003 ± 0.0015 \\
7 & 0.3331 ± 0.0670 & 0.0002 ± 0.0013 \\
8 & 0.4057 ± 0.0285 & 0.0221 ± 0.0092 \\
9 & 0.3970 ± 0.0367 & 0.0019 ± 0.0032 \\
\end{tabular}%
}
\caption{Comparison of actual vs. random community consistency scores (mean ± standard deviation). Each community is significantly different than the random score suggesting that the communities are consistent across subjects and runs.}
\label{tab:community_cscore_stats}
\end{table}

\section{Conclusion}

In this paper, we introduced a method for detecting overlapping communities that take the higher order dependencies in brain networks. While prior work in the area of brain connectivity focuses on detecting non-overlapping communities, our approach makes two key contributions. First,  we construct a hypergraph from fMRI data using sparse linear regression to capture the higher order dependencies between ROIs. Second, we apply spectral clustering to detect communities on the line graph of the hyperedges resulting in overlapping node communities. Finally, we learn the group-level community structure using the association matrix. Applying our framework to the HCP data yields symmetric and localized communities. These communities are mostly functional combinations of Yeo regions, and highly resemble previously discovered resting-state connectivity patterns. The communities also overlap, which shows connections between known areas of cognitive function. The consistency of the detected communities across subjects is also quantified. Future work will focus on improving the method with weighted hyperedges and selection of hyperparameters for the sparse regression.



\bibliographystyle{IEEEbib}
\bibliography{refs}

\end{document}